\documentclass[letterpaper, 10 pt, conference]{ieeeconf}  
\usepackage{amsmath,amsfonts}
\usepackage{algorithmic}
\usepackage{algorithm}
\usepackage[algo2e,ruled,vlined]{algorithm2e}
\usepackage{array}
\usepackage{textcomp}
\usepackage{stfloats}
\usepackage{url}
\usepackage{verbatim}
\usepackage{graphicx}
\usepackage{cite}
\usepackage{url} 
\usepackage{soul}
\usepackage{multirow}
\usepackage{nicematrix}
\usepackage{subfigure} 
\usepackage{xcolor}
\usepackage{colortbl}
\usepackage{bm}
\usepackage{textcomp}

\usepackage[hang,small,bf]{caption}
\usepackage[subrefformat=parens]{subcaption}
\usepackage{array, booktabs, multirow}
\usepackage{afterpage}
\usepackage{tabularray}
\usepackage{bbm}

\newcommand{\PreserveBackslash}[1]{\let\temp=\\#1\let\\=\temp}

\newcommand{\argmin}{\mathop \mathrm{arg~min}\limits}
\usepackage{mathcomp}

\usepackage{multicol}

\usepackage{tabularray}
\UseTblrLibrary{diagbox}
\usepackage{multicol}

\IEEEoverridecommandlockouts                             
\usepackage[utf8]{inputenc}
\usepackage[T1]{fontenc}

\title{\LARGE \bf
Digital Twin-Empowered Cooperative Autonomous Car-sharing Services: Proof-of-Concept
}

\author{Kazuma Nonomura,
Kui Wang,
Zongdian Li, 
Tao Yu, and 
Kei Sakaguchi
\thanks{*This work was supported in part by the National Institute of Information and Communications Technology (NICT) Japan-US Networking Opportunity (JUNO) Program under Grant \#22404.}
\thanks{K. Nonomura, K. Wang, Z. Li, T. Yu, and K. Sakaguchi are with the Institute of Science Tokyo, Japan (Email: \{nonomura, kuiw, lizd, yutao, sakaguchi\}@mobile.ee.titech.ac.jp).}%
}

\begin{document}

\maketitle
\thispagestyle{empty}
\pagestyle{empty}

\begin{abstract}

This paper presents a digital twin-empowered real-time optimal delivery system specifically validated through a proof-of-concept (PoC) demonstration of a real-world autonomous car-sharing service. This study integrates real-time data from roadside units (RSUs) and connected and autonomous vehicles (CAVs) within a digital twin of a campus environment to address the dynamic challenges of urban traffic. The proposed system leverages the Age of Information (AoI) metric to optimize vehicle routing by maintaining data freshness and dynamically adapting to real-time traffic conditions. Experimental results from the PoC demonstrate a 22\% improvement in delivery efficiency compared to conventional shortest-path methods that do not consider information freshness. Furthermore, digital twin-based simulation results demonstrate that this proposed system improves overall delivery efficiency by 12\% and effectively reduces the peak average AoI by 23\% compared to the conventional method, where each vehicle selects the shortest route without considering information freshness. This study confirms the practical feasibility of cooperative driving systems, highlighting their potential to enhance smart mobility solutions through scalable digital twin deployments in complex urban environments.

\end{abstract}


\section{Introduction}\label{sec:introduction}
The rapid growth of urban populations has increased the demand for efficient traffic management in major cities \cite{rudskoy2021digital}. Intelligent transportation systems (ITS) and connected autonomous vehicles (CAVs) are expected to transform urban mobility by enhancing safety, reducing congestion, and optimizing resource usage \cite{seth2019traffic}. Autonomous car-sharing systems, which utilize CAVs to provide on-demand mobility, are especially promising as they can significantly reduce private vehicle ownership \cite{hao2018shared}. Consequently, both public and private sectors have focused on developing autonomous car-sharing solutions to address urban transportation challenges \cite{javanshour2021performance}.

Despite such progress, current autonomous mobility services primarily focus on autonomous mobility on demand (AMoD) systems led by major technology companies. For instance, Waymo has deployed commercial driverless ride-hailing services in the U.S. \cite{Waymo}, and Baidu Apollo has launched robotaxi programs in Chinese cities \cite{zhou2023robotaxi}. These trials have shown that autonomous vehicles can function effectively in well-defined zones, providing user convenience and gathering data to enhance safety and performance. However, most existing systems prioritize local autonomy, relying on onboard sensors and limited vehicle-to-vehicle interaction. Global traffic coordination and optimization are underutilized due to the reliance on local sensing over cloud-based decision-making.

Digital twins offer a crucial solution to these limitations. A digital twin creates a virtual representation of physical systems, continuously updated with real-time data to reflect the environment’s current state. In smart mobility, it enables real-time monitoring, adaptive decision-making, and dynamic traffic optimization by mirroring real-world scenarios in cyberspace.

Without digital twin integration, car-sharing systems cannot achieve optimal traffic flow or collaboration among vehicles. To address this, we propose a digital twin-based cooperative car-sharing system. Our framework integrates real-time data from roadside units (RSUs) and CAVs through a vehicle-to-everything (V2X) network. This data feeds into a cloud-based digital twin, enabling adaptive decision-making to (i) match users with CAVs on demand and (ii) optimize cooperative route planning for improved travel efficiency. The system has been tested in a real-world campus environment as a proof-of-concept (PoC) demonstration, validating its effectiveness.
The main contributions of this paper are summarized as follows:

\begin{itemize}
    \item \emph{Digital Twin-based System Design}: We propose a smart mobility digital twin (SMDT)-based cooperative car-sharing framework that integrates real-time sensor data from RSUs and CAVs for global traffic monitoring and decision-making in the cloud.
    
    \item \emph{Cooperative Driving and Scheduling}: We develop a method for cooperative driving and CAV scheduling to optimize route planning, enabling efficient and flexible matching of users to vehicles, as well as holistic route optimization based on current traffic conditions.
    
    \item \emph{Implementation and PoC Demonstration}: The proposed system is validated throughout a real-world campus testbed and extensive digital twin-based simulations, showcasing the system’s feasibility and effectiveness through on-site experiments.

\end{itemize}

The rest of this paper is organized as follows. Section II reviews related works on autonomous car-sharing. Section III outlines the digital twin-based system architecture and formulates the scheduling and routing as an optimization problem. Section IV presents the system implementation and PoC experiments, followed by the conclusion and future work in Section V.

\section{Related Works}

Research on cooperative autonomous driving focuses on overcoming perception limitations through real-time information sharing among connected vehicles, RSUs, and cloud platforms. The 5G Automotive Association (5GAA) highlights the importance of hybrid control systems combining decentralized decision-making with cloud-based global optimization \cite{5gaa_roadmap}.

Digital twin technology has become essential for smart mobility. Studies such as \cite{its-dt} demonstrate its effectiveness in real-time pedestrian detection through cooperative perception between AVs and RSUs. However, most systems rely on either cloud computing or single-layer edge computing, leading to challenges in balancing large-scale data processing with low-latency requirements.

To address these challenges, researchers have explored hierarchical architectures that combine edge and cloud computing for real-time traffic optimization \cite{vissim_traffic_optimization}. While promising, practical deployment remains limited. Recent experiments \cite{wang2024smdt, wang2024augmented} have validated mobility digital twins, demonstrating their role in supporting autonomous navigation and adapting to dynamic environments. Dynamic digital twins, updated via wireless networks, depend on experience-based learning and synchronization between the digital and physical systems \cite{omar-dt}.

A major challenge for digital twins is information decay, where outdated data can degrade system performance. To overcome such an issue, Age of Information (AoI)-based methods have been proposed to maintain data freshness. For instance, \cite{aoi-uav-navigation} introduced an AoI-constrained reinforcement learning method for UAV networks, showing that fresh data improves system efficiency.

Despite progress, most studies rely on simulations or small-scale tests, lacking validation in real-world, large-scale settings. This paper addresses these gaps by introducing a mobility digital twin system with AoI-driven traffic graph networks. Our work enhances autonomous vehicle-based delivery efficiency through real-world validation and adaptive route optimization.

\section{System Design}\label{sec:system}
\begin{figure}[t]
\centerline{\includegraphics[keepaspectratio, width=0.5\textwidth]
{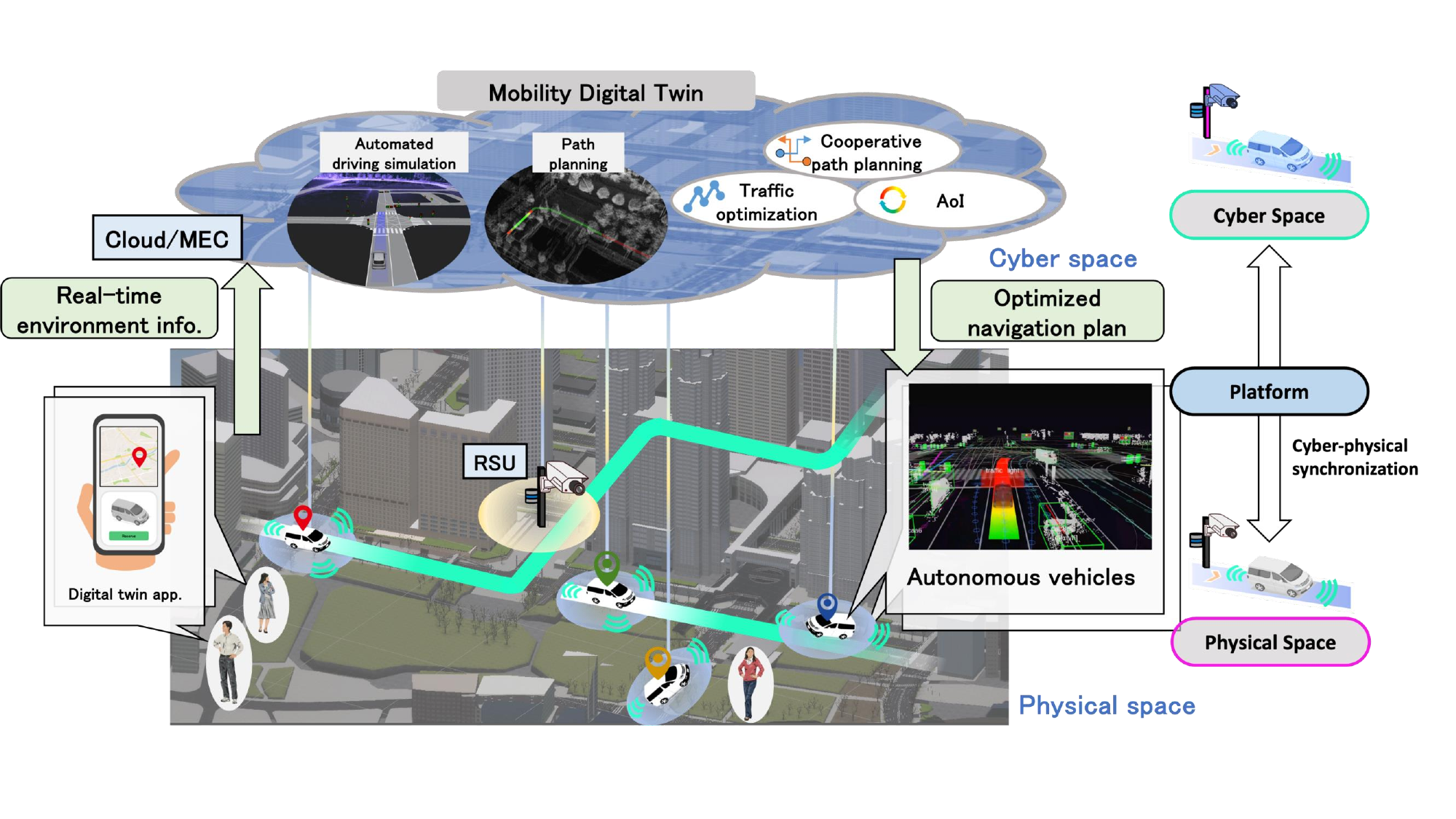}}
\caption{Digital twin-Empowered cooperative autonomous car-share services}
\label{fig:sys_overview}
\end{figure}

\begin{figure*}[t]
\centerline{\includegraphics[keepaspectratio, width=0.8\textwidth]{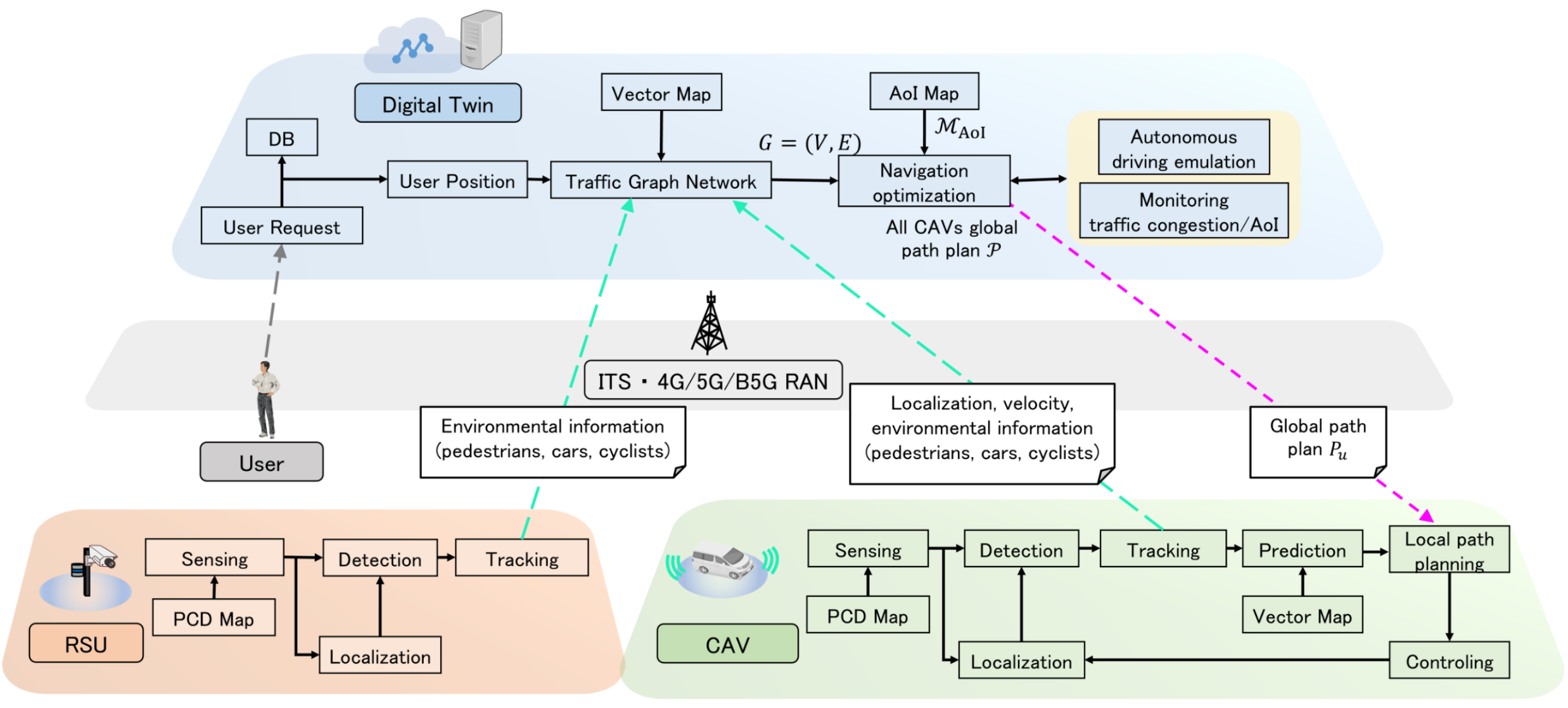}}
\caption{Proposed system architecture}
\label{fig:sys_arch}
\end{figure*}

The proposed cooperative autonomous carsharing system, which leverages autonomous vehicles, RSUs, and the mobility digital twin, is illustrated in Fig.~\ref{fig:sys_overview}.
The system integrates object recognition, digital twin technology, V2X communication, and route optimization. The functionalities of each component are detailed in Fig.~\ref{fig:sys_arch}. This system introduces several innovations by integrating a hierarchical digital twin framework with subsystems that include autonomous vehicles, RSUs, and mobile edge computing (MEC)/Cloud platforms. The design uniquely combines edge and cloud resources, ensuring both low-latency local perception and globally optimized traffic management.

\subsection{System Architecture}
RSUs are equipped with LiDAR sensors, cameras, edge computing units, and V2X communication devices. The RSUs and autonomous vehicles collaboratively perform object detection and environmental observation, thereby creating localized mobility digital twin models within the cyber-physical space.

The perception data gathered by RSUs and CAVs is then continuously transmitted to the MEC/Cloud, where the large-scale mobility digital twin is dynamically updated in real time. This large-scale digital twin leverages object detection results collected from edge devices to perform traffic analysis and optimize routing plans for multiple autonomous vehicles. The optimized delivery routes are subsequently fed back to individual vehicles via vehicle-to-network (V2N) communication.

A key innovation of the proposed system is its ability to go beyond static route optimization by directing autonomous vehicles to explore areas not covered by RSUs. This exploration actively gathers additional data, enabling the system to fill information gaps and maintain a city-scale, quasi-dynamic mobility map within the digital twin. By balancing exploration and exploitation, the system enhances comprehensive traffic monitoring and adaptive route optimization, providing a scalable solution for future smart mobility deployments.

\subsection{Optimization}

This study assumes an environment with both fixed sensors (e.g., RSUs) and mobile sensors (e.g., sensors on autonomous vehicles) coexisting in a cooperative mobility system involving multiple autonomous vehicles. RSUs and vehicles continuously transmit sensor data to the server, where the digital twin reflects real-time traffic conditions. A critical challenge is maintaining data freshness by deciding when outdated information should be discarded. Fixed sensors, such as RSUs, consistently update real-time data, while mobile sensor data gradually loses value as vehicles move. To address this, the system encourages exploring actions that maintain a consistent level of information freshness for effective optimization.

\textbf{Problem settings}

Let $\mathcal{U}$ be the set of autonomous vehicles in the system. The road network is represented as a directed graph $G = (V, E)$, where $E$ denotes road segments and $V$ represents intersections or merging points. The travel time along an edge $e^{i,j} \in E$ is given by $T^{i,j}(t)$. At time $t$, the planned route of vehicle $u \in \mathcal{U}$, starting at position $p_u \in V$, is defined as $P_u=\left\{p_{u(1)},\cdots,p_{u(T)}\right\}\in V^T$.

The optimization problem is divided into two phases:
\begin{enumerate}
\item Empty Vehicle Phase $(p_u \Rightarrow s_u)$: The vehicle travels to the pickup location $s_u$ while exploring areas with low information freshness.
\item User Transport Phase $(s_u \Rightarrow g_u)$: The vehicle transports the user to the destination $g_u$, prioritizing the fastest available route.
\end{enumerate}

\textbf{Navigation Optimization Method}

A) Empty Vehicle Phase $(p_u\Rightarrow s_u)$

During this phase, the vehicle minimizes the following cost function to optimize its route while collecting fresh data:

\begin{equation}
\argmin_{P_u}\sum_{\tau=1}^{T-1} T^{p_{u(\tau)},p_{u(\tau+1)}}{(t)} 
\end{equation}
s.t.
\begin{equation*}
\begin{split}
    &p_{u(1)} = p_u , p_{u(T)} = s_u\\
    &{p_{u(\tau)}}\bigcap \left(\{P_u\}\setminus p_{u(\tau)}\right)=\emptyset, \forall p_{u(\tau)}\in P_u\\    &\beta,v_\text{traffic},k_\text{max} \in \mathbb{R}_+\\
\end{split}
\end{equation*}

The travel time $T^{i,j}(t)$ along edge $e^{i,j}$ accounts for dynamic conditions and baseline travel times is modeled as follows by integrating AoI and Fundamental Diagram (FD) based on the Kinematic-Wave theory, where $\Delta_m$ represents AoI and $k(x,t)$ represents the real-time traffic density.
\begin{equation}
T^{i,j}(t) = \big(T^{i,j}_{\text{dynamic}}(t) - T^{i,j}_{\text{free}}\big) e^{-\beta \Delta_m(t)} + T^{i,j}_{\text{free}}
\end{equation}
The dynamic travel time is defined as $T^{i,j}_{\text{dynamic}}(t) = \frac{l}{v_{\text{traffic}}(t)} = \frac{l}{v_{\text{free}}\left(1 - \frac{k(x,t)}{k_{\text{max}}}\right)}$. $T^{i,j}_\text{free}$ is the travel time under free-flow conditions, given by $T^{i,j}_\text{free}=\frac{l}{v_\text{free}}$. $\Delta_m(t)$ is the AoI for edge $e^{i,j}$, given by $\Delta_m(t) = t - u^\prime_m$. $T_\text{dynamic}(t)$ represents the latest real-time traffic conditions stored in the digital twin. And $T_\text{free}$ represents baseline travel time obtained from a static map without real-time data, as described in \cite{travel-time2} by $q(x,t)=k(x,t)v(x,t)$, $v_\text{traffic}(x,t)=v_\text{free}\left(1-\frac{k(x,t)}{k_\text{max}}\right)$. Using the computed travel time $T^{i,j}(t)$, we construct a weight matrix $T_\text{AoI}$ that accounts for information freshness decay: $\bm{T}_\text{AoI}(t) = \{T^{i,j}(t, \Delta_{i,j})\}_{i,j \in V}$. By using this weight matrix, the system optimizes the exploratory behavior of unoccupied autonomous vehicles while balancing travel time and data freshness.

B) User Transport Phase$(s_u\Rightarrow g_u)$

When carrying passengers, the vehicle minimizes travel time by using the latest traffic data. This means setting $\beta=0$, so that $T^{i,j}(t) = T_\text{dynamic}(t)$. Thus, the vehicle strictly follows the fastest available route without prioritizing exploration.

By solving this optimization problem, the proposed system allows autonomous vehicles to balance exploration and exploitation:
\begin{itemize}
    \item When unoccupied, vehicles explore areas with low information freshness, updating the digital twin while still ensuring reasonable pickup times.
    \item When carrying passengers, vehicles prioritize the fastest travel time, ensuring high-quality transportation service.
\end{itemize}
This adaptive routing strategy effectively maintains real-time data freshness while optimizing passenger delivery efficiency.

\section{Proof-of-Concept on Digital Twin-based Autonomous Driving}\label{sec:poc}

\begin{table*}[t]
\centering
\setlength{\tabcolsep}{8pt} 
\caption{Hardware for Autonomous Car-sharing Digital Twin}
\label{tab:mobility_digital_twin_hardware}
\begin{tabular}{l|l p{0.45\linewidth}c}
\toprule
\textbf{Type} & \textbf{Device Name} & \textbf{Specifications} & \textbf{Amount} \\ 
\midrule
\multirow{2}{*}{Vehicles} 
    & ZMP RoboCar MV2 OBU & Autonomous vehicle with driving controller (1 seater) & 1 \\
    & Toyota MiniVan Estima OBU & Autonomous vehicle with driving controller (6 seater) & 1 \\ 
\midrule
\multirow{2}{*}{Sensors} 
    & RS-LiDAR-32 & Position: Vehicle edges, Range: 200 m, Accuracy: $\pm$ 3 cm, Rotation speed: 10/20 Hz & 2 \\
    & RS-LiDAR-80 & Position: RSU edge, Range: 230 m, Accuracy: $\pm$ 3 cm, Rotation speed: 5/10/20 Hz & 3 \\ 
\midrule
\multirow{2}{*}{Car-share Application} 
    & Apple iPad Pro 12.9 (5th) & Position: User, Function: Reservation \& Remote monitoring  & 2 \\
    & Lenovo Thinkpad X1 Carbon & Position: Car-sahre provider, Function: Remote monitoring & 1 \\
\midrule
\multirow{1}{*}{Communication} 
    & WiMAX NIC & Position: Vehicle edges, Downlink: 120 Mbit/s, Uplink: 60 Mbit/s, Maximum coverage: 30 miles & 2 \\
\midrule
\multirow{3}{*}{Cloud \& Edge Servers} 
    & Jetson AGX Orin & Position: RSU edges, OS: Ubuntu 20.04 (JetPack 5.0.2), ROS: Galactic, Autoware.Universe & 3 \\
    & Autoware PC & Position: Vehicle edge,  OS: Ubuntu 22.04, ROS: Humble, Autoware.Universe & 1 \\
    & Digital Twin Engine & Position: Cloud, OS: Ubuntu 22.04, ROS: Humble, AWSIM & 1 \\ 
\bottomrule
\end{tabular}
\end{table*}

\subsection{Implementation}\label{subsec:implementation}
\begin{figure}[t]
\centering
    \begin{tabular}{c}
        \begin{tabular}{cc}
            \centerline{\includegraphics[keepaspectratio, width=0.4\textwidth]
            {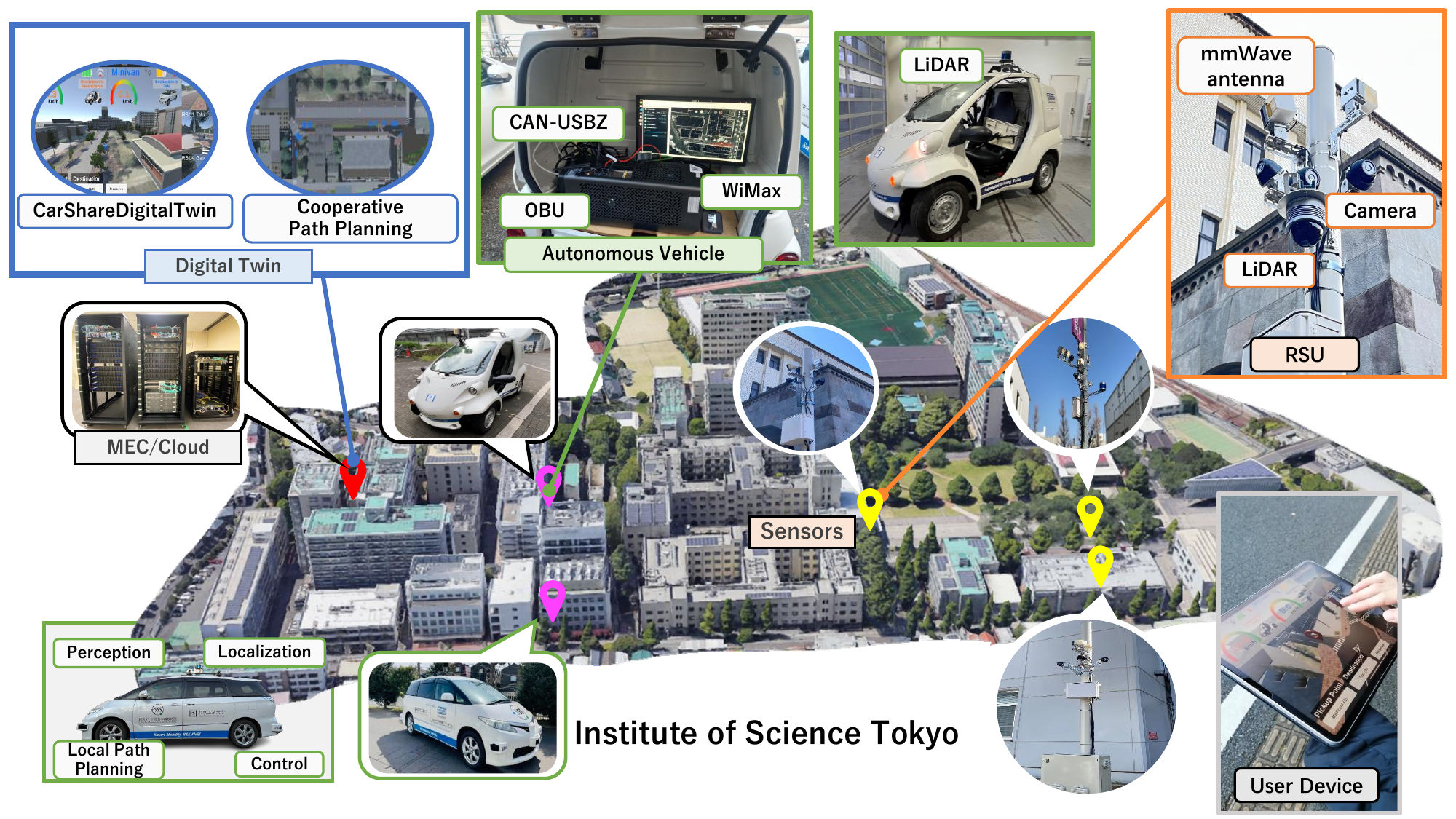}}
            \\
            \small (a) Experimental Field
            \\\\
            \centerline{\includegraphics[keepaspectratio, width=0.45\textwidth]
            {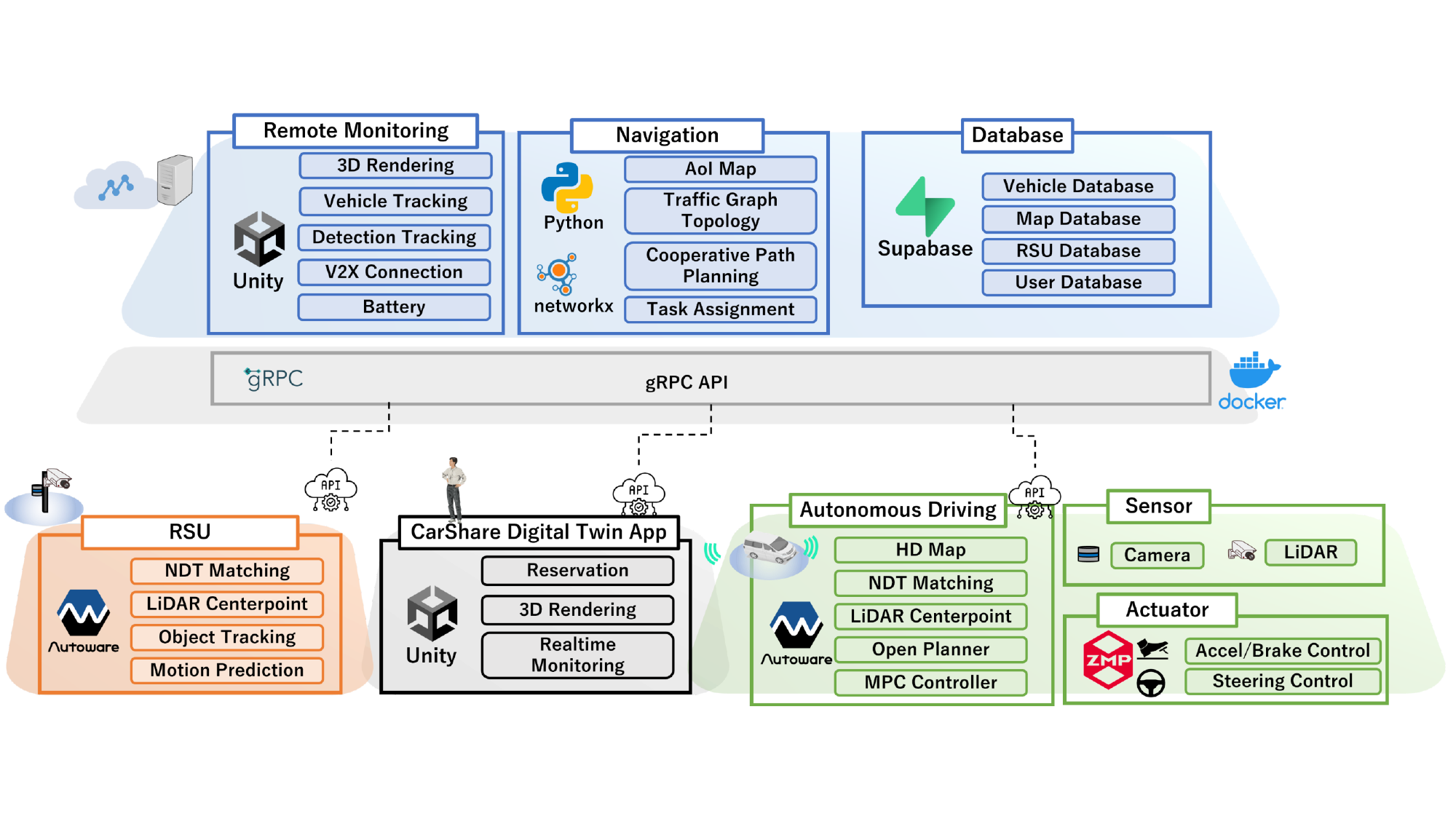}}
            \\
            \small (b) Software Configuration
        \end{tabular}
    \end{tabular}
    \caption{Hardware/Software Configuration}
    \label{fig:hw_sw}
\end{figure}

The smart mobility field for the proof of concept is placed at Institute of Science Tokyo, Ookayama campus, Japan.
Fig.\ref{fig:hw_sw} shows the overall hardware and software configuration.
The field includes two autonomous vehicles, three RSUs, one digital twin server, and two user devices.

We have used two autonomous vehicles: a one-seater RoboCar MV2\cite{MV2} and a six-seater MiniVan.
On the rooftop of each vehicle, a 32-layer 3D LiDAR (RS-LiDAR 32\cite{robosense}) is mounted and connected to the OBU (On-Board Unit). The OBU ran Ubuntu22.04 as its operating system, with ROS2 (Robot Operating System 2) Humble and Autoware.Universe installed.
These were used to perform self-localization and object recognition (including object detection and multi-object tracking) using point cloud data.
Additionally, the Autoware PC processed the self-localization and surrounding environmental data obtained from point cloud processing. This enabled it to compute route planning to the destination and generate a collision avoidance trajectory for obstacles.
The computed target trajectory was then converted into control commands for actuators, such as steering angles, acceleration, and braking. Communication with the Electronic Control Unit (ECU) of the actuators was handled via a CANUSB-Z\cite{CANUSB-Z} interface, enabling data exchange over the CAN bus.

For the RSUs, each was equipped with a fixed 80-layer 3D LiDAR (RS-LiDAR 80). In addition, each RSU had one NVIDIA Jetson installed as an edge computing resource to process LiDAR point cloud data locally.
Specifically, the RSUs performed LiDAR-based object detection and tracking, streaming the processed data to the digital twin server. This enabled real-time object-level mapping from the physical space to the virtual space.

\subsection{Digital Twin-based Practical Car-Sharing Service}\label{subsec:poc1}
Next, to verify the feasibility of the proposed system, a real-world proof-of-concept (PoC) experiment was conducted using actual autonomous vehicles.
A practical implementation test of the autonomous driving car-sharing service was conducted at the Ōokayama Campus of Tokyo Science University, involving actual car-sharing users. The PoC1 scenario overview is shown in Fig. \ref{fig:poc_overview_carshare_service}. And the detailed steps of the PoC experiment are as follows:

\begin{figure}[t]
\centerline{\includegraphics[keepaspectratio, width=0.45\textwidth]
{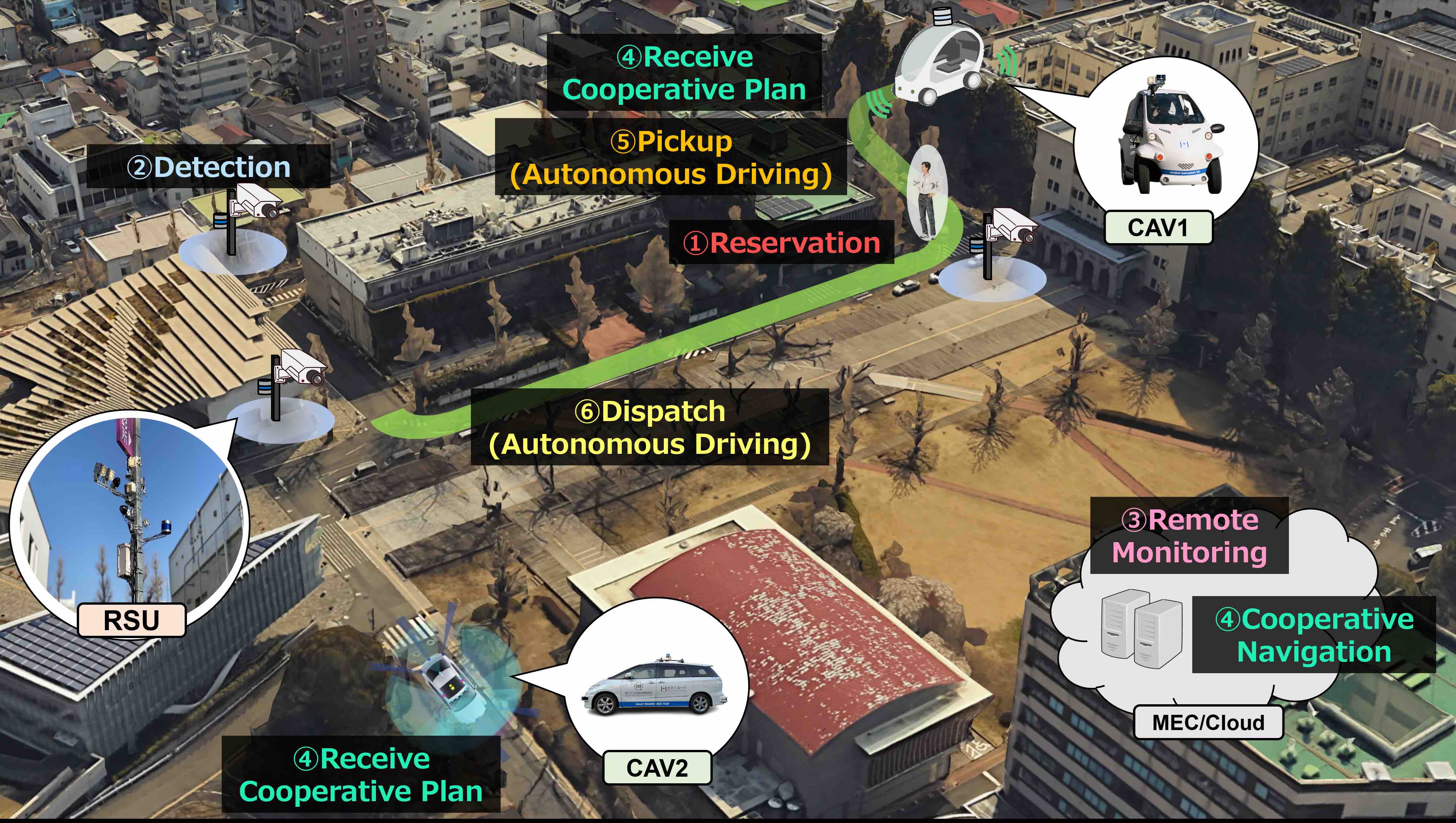}}
\caption{PoC1 Scenario: Mobility Digital Twin-based  CarShare Service}
\label{fig:poc_overview_carshare_service}
\end{figure}

\begin{enumerate}
\item Multiple users used the autonomous car-sharing service.
\item The Mobility Digital Twin (DT) collected sensing data from both RSUs and autonomous vehicles.
\item The real-world campus environment was replicated in real time within the digital twin.
\item Traffic congestion were monitored using real-time data.
\item Optimized route was planed to avoid congestion.
\item The optimized routes (checkpoints) were transmitted to the autonomous vehicles.
\item Autonomous vehicles followed the received optimal route to pick up users and then transported them to their destinations without human intervention.
\end{enumerate}

In practice, users could successfully reserve a vehicle through the car-sharing digital twin application, and the autonomous vehicle arrived at the designated pick-up location.
Although manual assistance was required for passengers to board, the vehicle was able to autonomously transport users to their destinations after pickup, demonstrating the feasibility of the proposed system.

\begin{figure}[h]
\centerline{\includegraphics[keepaspectratio, width=0.45\textwidth]
{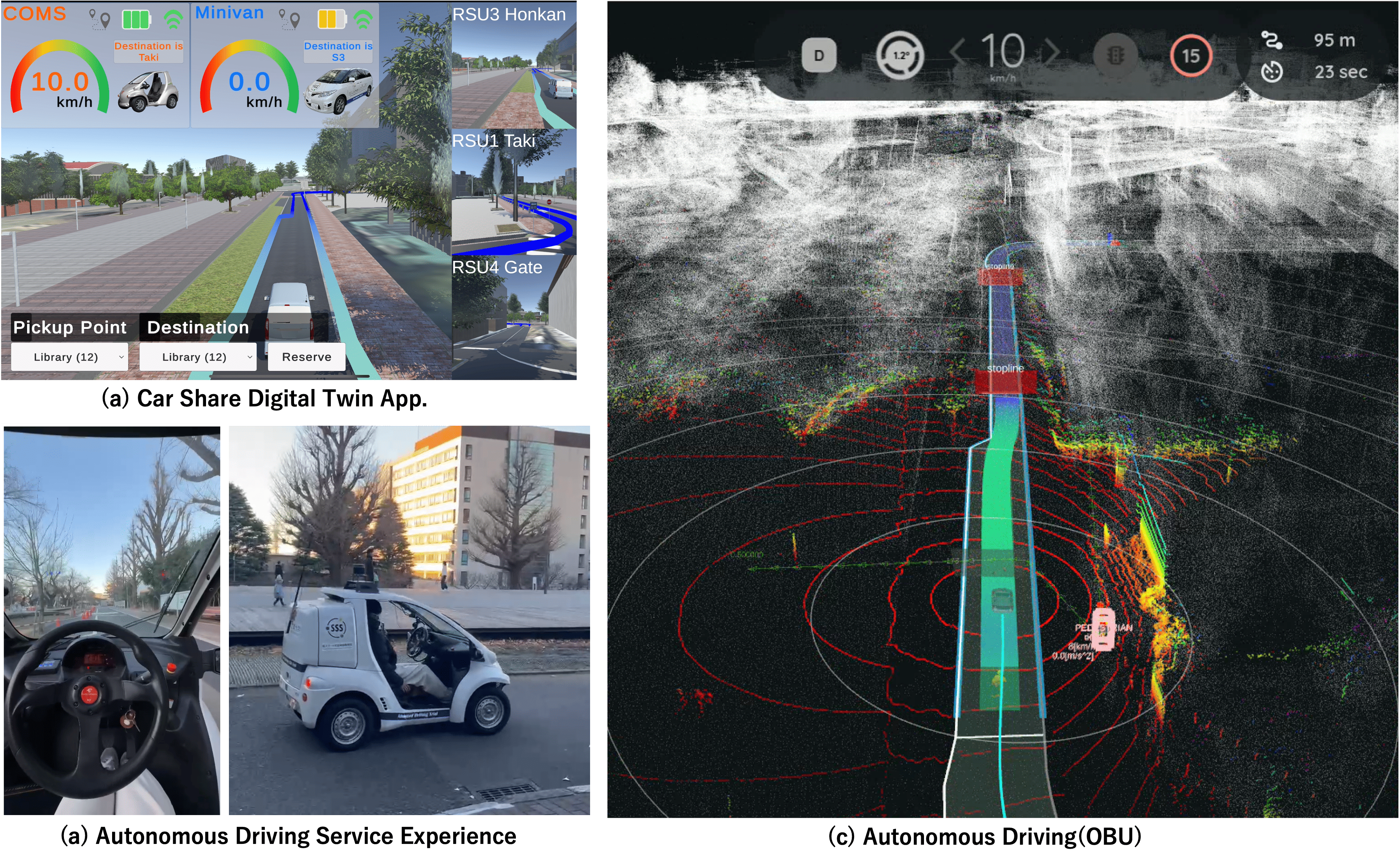}}
\caption{Autonomous Car-sharing Digital Twin Service}
\label{fig:cooperative_path_planning}
\end{figure}

\subsection{Cooperative Path Planning Evaluation}\label{subsec:poc2}

\begin{figure}[t]
\centerline{\includegraphics[keepaspectratio, width=0.45\textwidth]
{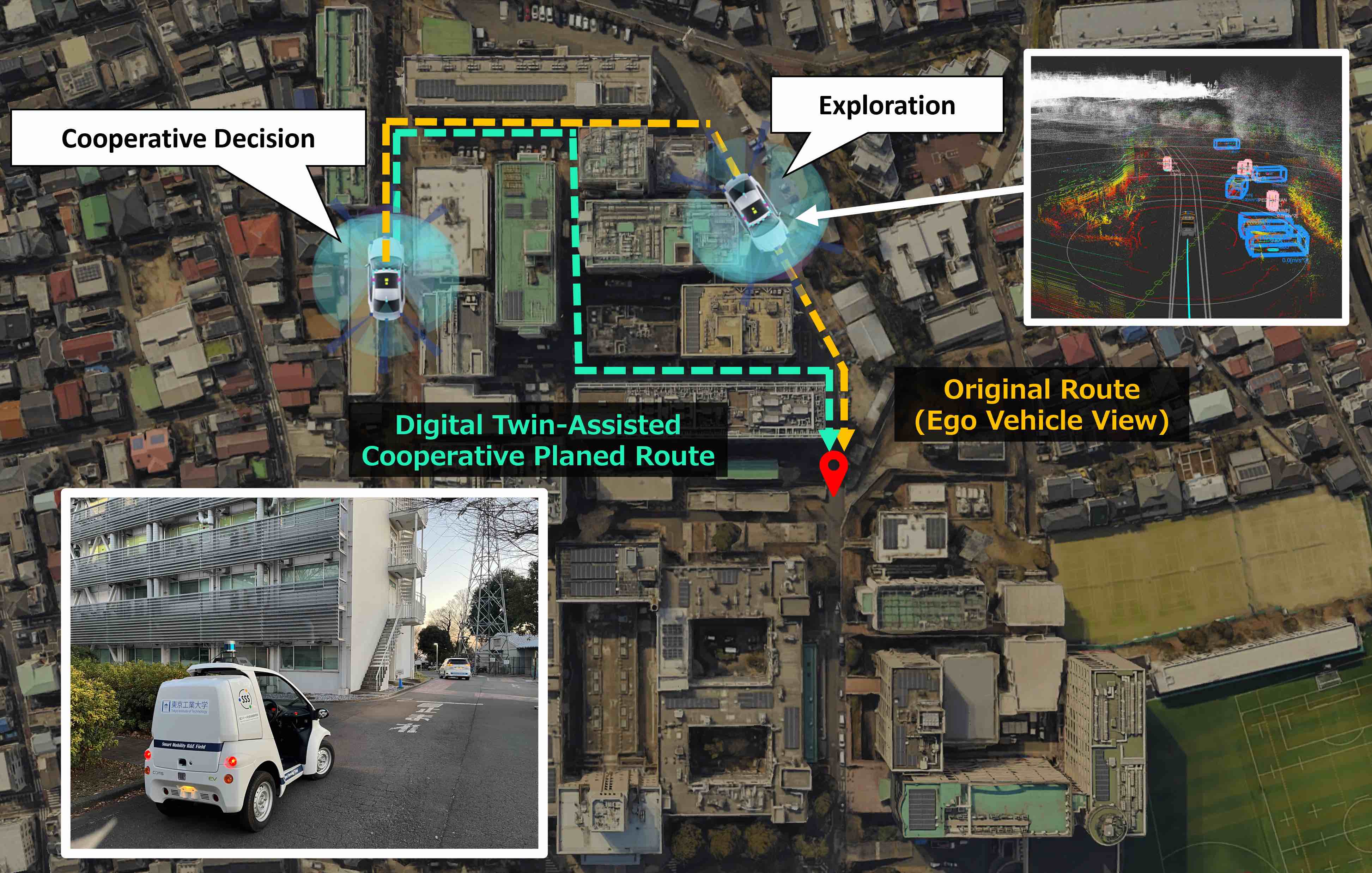}}
\caption{PoC2 Scenario: Cooperative Path Planning}
\label{fig:poc_overview_cooperative_path_planning}
\end{figure}

An outdoor evaluation experiment was conducted using two autonomous vehicles.
Due to limitations on the number of available vehicles, the experiment was performed with two CAVs. To ensure fair performance comparisons, the delivery time per task was measured multiple times under identical conditions, and the average was taken.
The experiment was conducted under normal campus conditions, where pedestrians and other vehicles were present and moving as usual.

\begin{enumerate}
    \item Two autonomous vehicles were prepared.
    \item The first CAV (forward vehicle) followed the yellow path in Fig.\ref{fig:poc_overview_cooperative_path_planning}.
    \item The second CAV (rear vehicle) started from its initial position (as shown in Fig. \ref{fig:poc_overview_cooperative_path_planning}) and transported the user autonomously to the destination marked by the red pin.
    \item The CAVs used onboard sensors for object recognition and shared real-time environmental data via V2N (Vehicle-to-Network) communication
    \item The rear autonomous vehicle planned its route based on two decision-making strategies:
    \begin{enumerate}
        \item Ego-Vehicle Decision (Conventional Approach): the rear CAV followed the same shortest path (yellow route) as the front CAV, without considering dynamic environmental updates.
        \item Cooperative Decision (Proposed Method): the rear CAV adapted its route based on real-time object detection from the front vehicle and selected the blue route to avoid traffic congestion.
    \end{enumerate}
\end{enumerate}
This experiment tested the effectiveness of cooperative mobility by comparing traditional shortest-path selection with adaptive decision-making using shared perception data.

Additionally, the comparison between ego-vehicle route planning and cooperative route planning using front vehicle detection data is shown in Table \ref{tab:navigation_time_comparison}. By leveraging object detection results from the front autonomous vehicle, the cooperative route planning approach achieved a 22.3\% reduction in delivery time compared to the ego-vehicle route planning approach. These results demonstrate that cooperative mobility strategies significantly improve efficiency by optimizing path selection based on shared real-time perception data.

\begin{table}
\caption{Comparison of Average Travel Time Between Cooperative and Ego-Vehicle Route Planning in the Proof-of-Concept Experiment}\label{tab:navigation_time_comparison}
\centering
\begin{tabular}{>{\hspace{0pt}}m{0.479\linewidth}|>{\hspace{0pt}}m{0.413\linewidth}} 
\hline\hline
Path Planning Method          & Average Travel Time[s]        \\ 
\hline
Cooperative Route Planning    & \textbf{146.35}  \\ 
\hline
Ego-Vehicle Route Planning & 198.95           \\
\hline\hline
\end{tabular}
\end{table}

\begin{figure}[!ht]
    \begin{tabular}{c}
        \begin{tabular}{cc}
            \centerline{\includegraphics[keepaspectratio, width=0.45\textwidth]
            {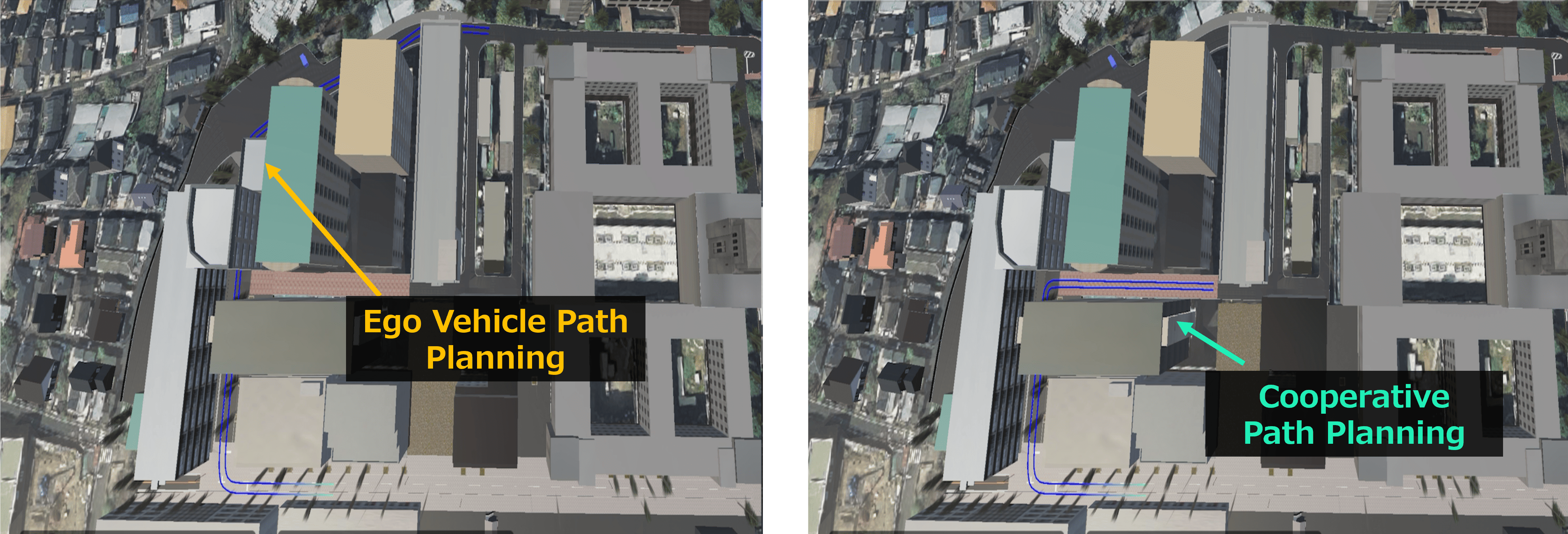}}
            \\
            \small (a) Car-sharing Digital Twin View
            \\
            \centerline{\includegraphics[keepaspectratio, width=0.45\textwidth]
            {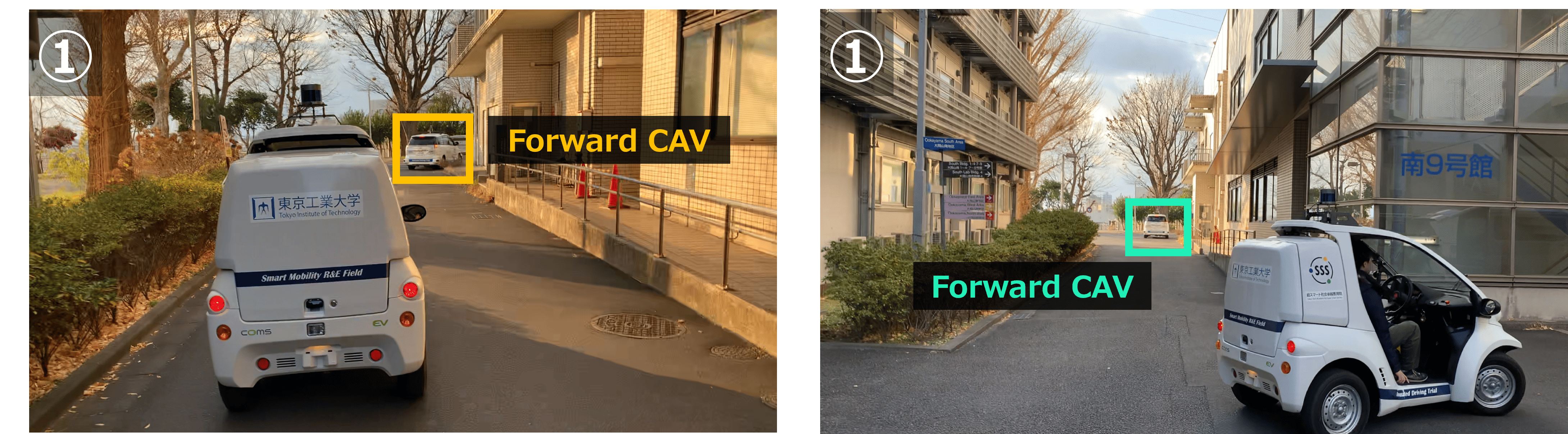}}
            \\
            \small (b) Route Switching at an Intersection (AV)
            \\
            \centerline{\includegraphics[keepaspectratio, width=0.45\textwidth]
            {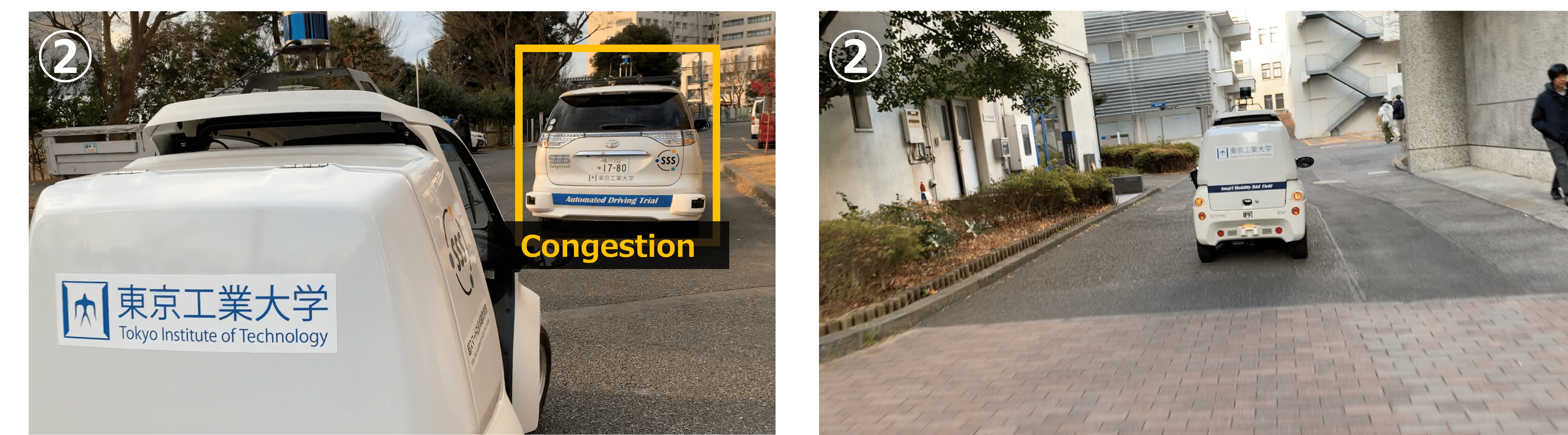}}
            \\
            \small (c) Route Switching beyond the Intersection (AV)
            \\
            \centerline{\includegraphics[keepaspectratio, width=0.45\textwidth]
            {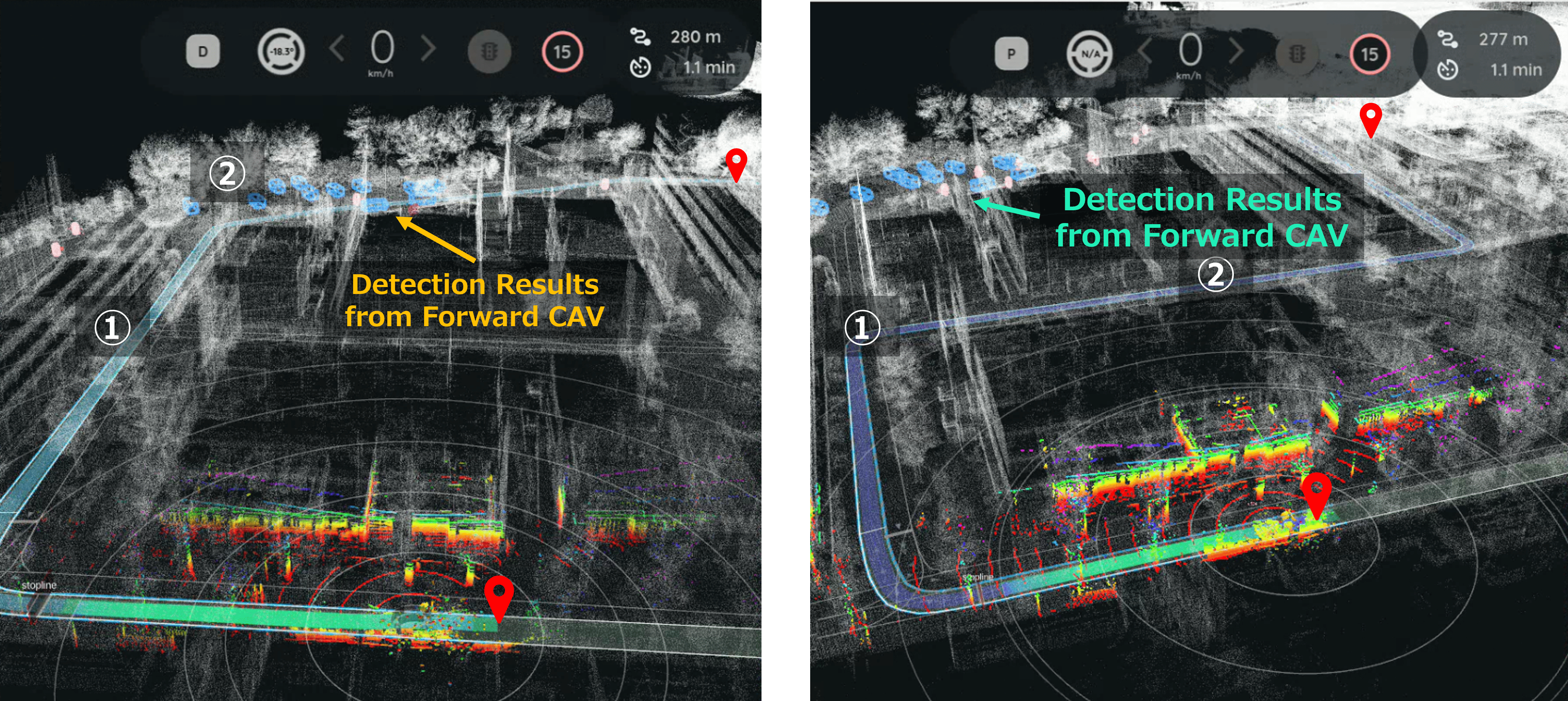}}
            \\
            \small (d) Autonomous Vehicle View
        \end{tabular} 
    \end{tabular}
    \caption{Ego Vehicle Planning (left) v.s. Cooperative Planning (right)}
    \label{fig:cooperative_decision}
\end{figure}

\subsection{Digital Twin-based Simulation}

We conducted a rule-based traffic simulation using AWSIM\cite{AWSIM}, ensuring full compliance with all Japanese traffic regulations. For general vehicle movement, we utilized AWSIM's RandomTrafficSimulator\cite{random_traffic_simulator}. This simulator determines vehicle behavior at intersections and lane branches defined in a vector map, where vehicles randomly select a lane from a list of possible next lanes. To introduce behavioral variations, we altered the random seed, allowing for diverse turning, merging, and straight-driving patterns in the simulation.

\begin{figure}[t]
\centerline{\includegraphics[keepaspectratio, width=0.45\textwidth]
{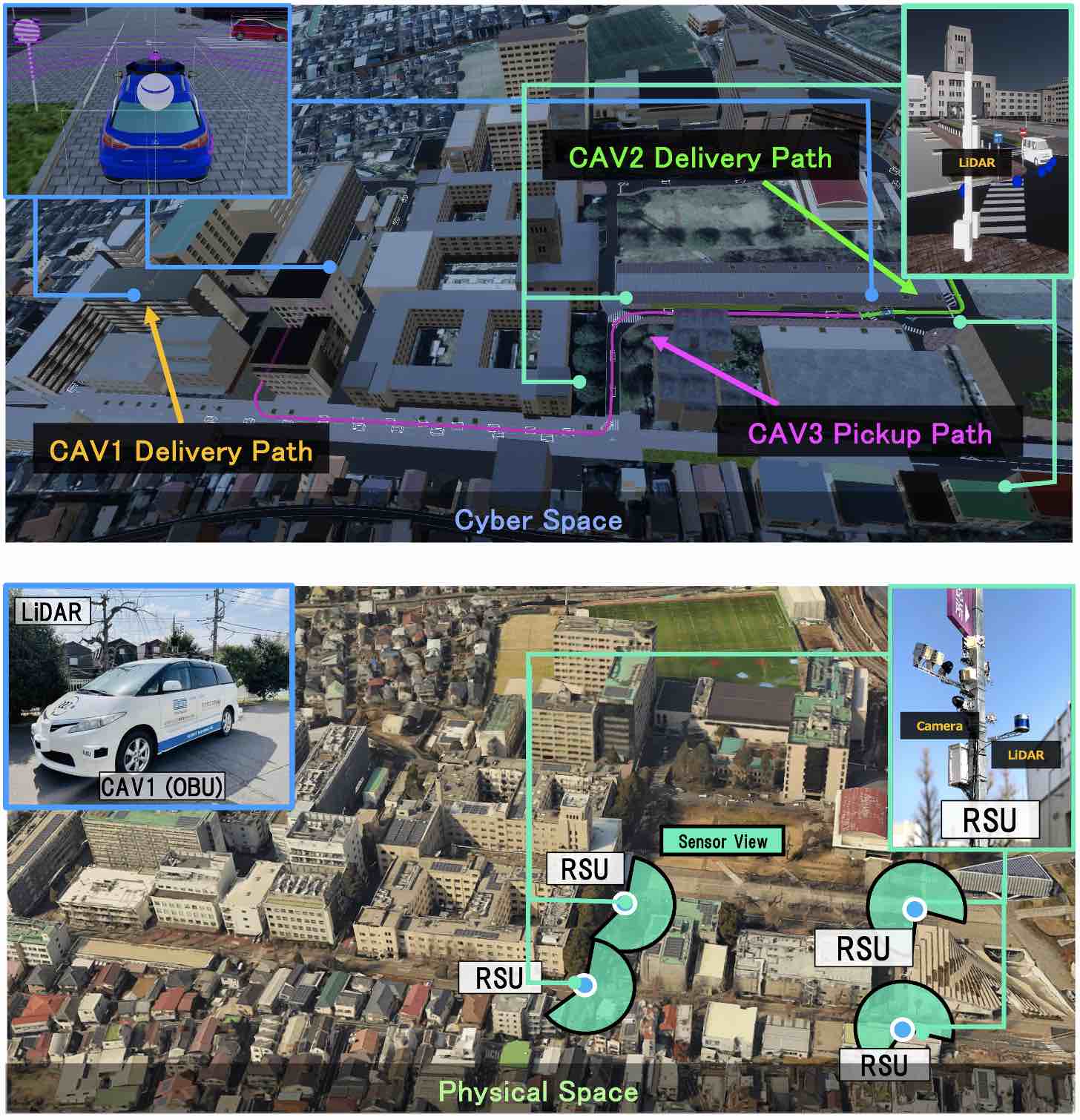}}
\caption{Car-sharing Digital Twin Simulation}
\label{fig:mobility_digital_twin_simulation}
\end{figure}

Based on the detected traffic objects' positions, the traffic density $k$ in the graph $G~(V,E)$ is estimated in real time.
To account for links where traffic density is not updated, we incorporate the AoI $\Delta_m$ into the traffic graph $G$ and compute the weighted adjacency matrix. The topology management of this traffic network was implemented using networkx\cite{networkx}.

Furthermore, a single task is defined as the process from when an autonomous vehicle starts moving toward a user’s location until the user is picked up and transported to the destination. The completion time of each task was measured for three autonomous vehicles. We conducted a simulation that repeatedly executed multiple user tasks, replicating the conditions of the previous subsection. In this simulation, the traffic volume was diversified by changing the seed for the action selection (turning right, left, or going straight) of general vehicles five times.

At time $t$, the spatially averaged Age of Information (AoI) in the digital twin can be expressed as follows:

\begin{equation}
\hat{\Delta}\mathcal{M}=\frac{1}{|\mathcal{M}|}\sum_{m\in\mathcal{M}} \Delta_m(t)
\end{equation}

This average AoI represents how outdated the information from vehicle-mounted sensors and RSU sensors reflected in the digital twin is compared to the real-world situation. In this study, the spatially averaged AoI is treated as a key system metric for evaluating the synchronization between the mobility digital twin and real-world conditions.

\subsection{Digital Twin Simulation Results}\label{subsec:simulation}
\begin{figure}[t]
\centerline{\includegraphics[keepaspectratio, width=0.45\textwidth]
{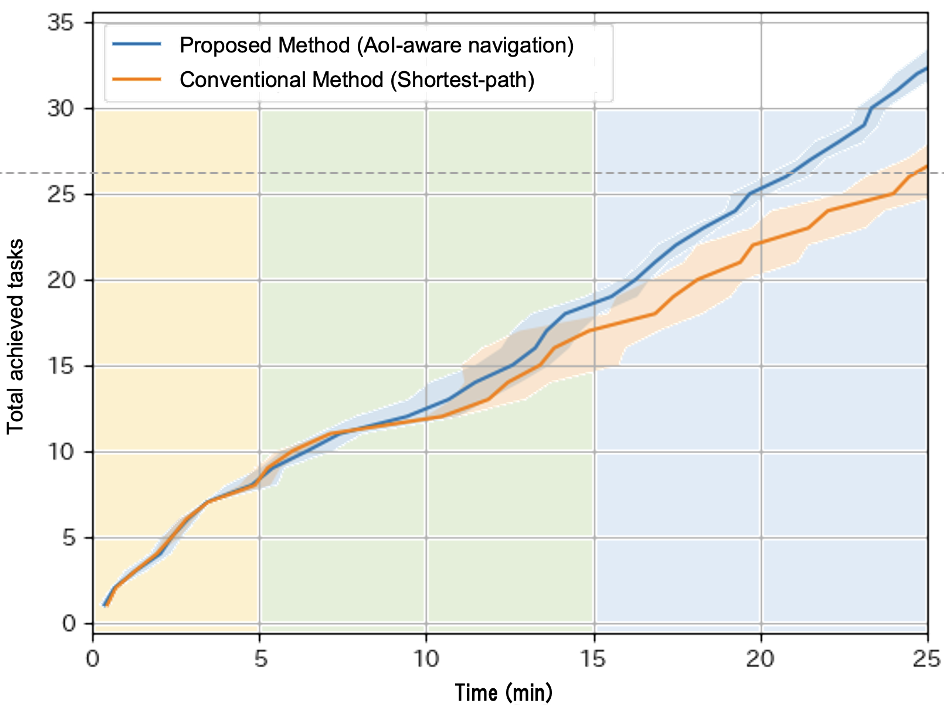}}
\caption{Total Delivery Tasks}
\label{fig:total_delivery_tasks}
\end{figure}

\begin{table}[]
\centering
\scriptsize
\renewcommand{\arraystretch}{0.5} 
\caption{Delivery Efficeincy in Different Traffic Conditions}
\label{tab:result_delivery}
\small
\begin{tblr}{
  colspec = {Q[c,1.6cm]Q[c,1.5cm]Q[c,1.5cm]Q[c,2.5cm]}, 
  cells = {c, },
  stretch = 0.9, 
  cell{1}{1} = {r=2}{},
  vlines,
  hline{1,3-5} = {-}{},
}
Conditions (min)   & Normal & Congestion & Post-congestion\\
      & $0\leq t\leq 5$ & $5\leq t\leq 15$ & $15\leq t$ \\
Proposal & \textbf{0.53 tasks/min} & \textbf{0.33 tasks/min} & \textbf{0.48 tasks/min}\\
Conventional & \textbf{0.53 tasks/min} & 0.30 tasks/min & 0.32 tasks/min
\end{tblr}
\end{table}

From Fig. \ref{fig:total_delivery_tasks}, it is shown that the proposed method, which incorporates AoI-aware route planning, completes delivery tasks faster than the conventional method, which does not consider AoI. 
The solid line indicates the expected delivery task completion time, while the shaded region denotes the confidence interval, both calculated over five different random seeds.
The total number of completed delivery tasks improved by 11.8\%.

\begin{figure}[t]
\centerline{\includegraphics[keepaspectratio, width=0.45\textwidth]
{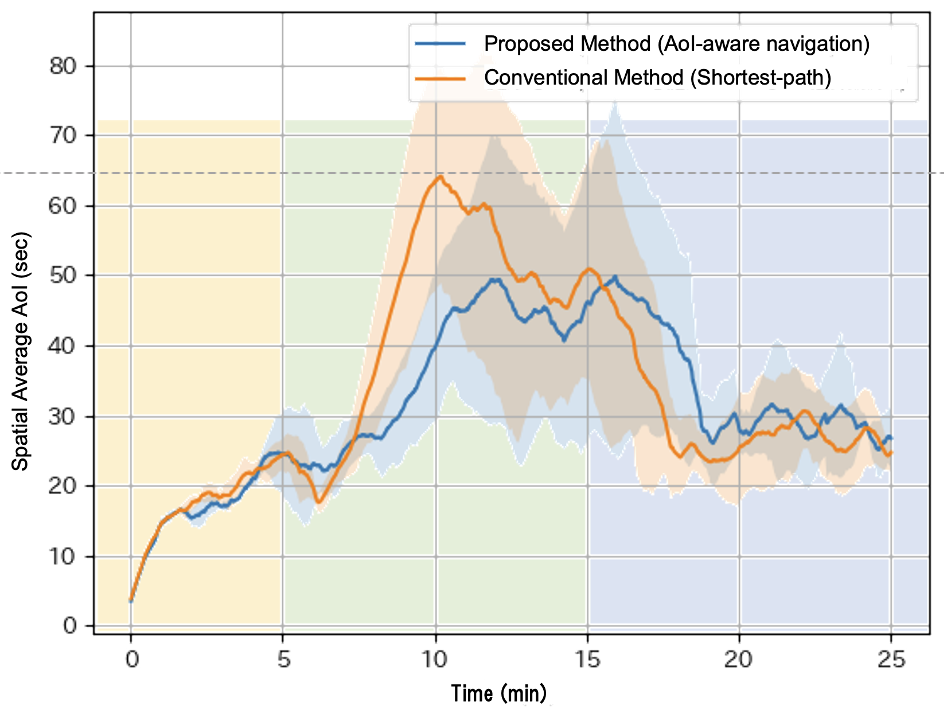}}
\caption{Spatial Average AoI}
\label{fig:spatial_average_aoi}
\end{figure}

\begin{table}[t]
\centering
\caption{Peak Spatial Average AoI}
\label{tab:result_peakaoi}
\small
\begin{tblr}{
  colspec = {Q[c,1.6cm]Q[c,1.5cm]Q[c,1.5cm]Q[c,2.5cm]}, 
  cells = {c},
  cell{1}{1} = {r=2}{},
  vlines,
  hline{1,3-5} = {-}{},
}
Conditions (min)   & Normal & Congestion & Post-congestion\\
      & $0\leq t\leq 5$ & $5\leq t\leq 15$ & $15\leq t $ \\
Proposal & 24.7sec  & \textbf{49.4sec} & \textbf{49.8sec}\\
Conventional & \textbf{24.4sec}  & 64.2sec & 50.9sec
\end{tblr}
\end{table}

Additionally, Fig. \ref{fig:spatial_average_aoi} and Table \ref{tab:result_peakaoi} present the expected value, confidence interval, and peak value of the spatially averaged AoI. From these results, it is evident that the proposed method reduces the peak value of the average AoI by 23.0\% compared to the conventional method. This indicates that the proposed approach successfully plans routes that maintain consistent information freshness across the mobility digital twin by leveraging multiple autonomous vehicles. Based on these findings, we confirm that the proposed method is resilient to unforeseen traffic congestion and maintains delivery efficiency without significant performance degradation.

\section{Conclusion and Future Works}\label{sec:conclusion}
This study proposed a cooperative mobility system for multiple autonomous vehicles using digital twin technology and demonstrated its effectiveness. By integrating hierarchical digital twins with edge computing and cloud environments, real-time traffic data acquisition and efficient route planning were achieved.
A real-world validation at Ōokayama Campus confirmed the feasibility of the proposed system through simulations and field experiments, including an autonomous car-sharing service. These results highlight the potential of digital twins in optimizing cooperative autonomous driving service operations for future smart mobility solutions.

\bibliographystyle{IEEEtran.bst}
\bibliography{bibliography}

\end{document}